\documentclass[prb,preprint]{revtex4-1} 

\usepackage{amsmath}  
\usepackage{amsfonts} 
\usepackage{graphicx} 

\begin{document}


\title{Teaching labs for blind students: equipment to measure the inertia of simple objects}

\author{A. Lisboa}
\email{alfredo.navarro@usm.cl} 
\affiliation{Departamento de F\'isica, Universidad T\'ecnica Federico Santa Mar\'ia, Av. Espa\~na 1680, Valpara\'iso 11520, Chile} 
\author{F. J. Peña} 
\email{francisco.penar@usm.cl}
\affiliation{Departamento de F\'isica, Universidad T\'ecnica Federico Santa Mar\'ia, Av. Espa\~na 1680, Valpara\'iso 11520, Chile}
\affiliation{Millennium Nucleus in NanoBioPhysics (NNBP), Av. Espa\~na 1680, Valpara\'iso 11520, Chile;}



\date{\today}

\begin{abstract}
This article explains and illustrates the design of a laboratory experience for blind students to measure the inertia of simple objects, in this case, that of a disc around its axis of symmetry. Our adaptation consisted in modifying the data collection process, where we used an open-source electronic platform to convert visual signals into acoustic signals. This allows one of the blind students at our University to participate simultaneously as their classmates in the laboratory session corresponding to the mechanics unit of a standard engineering course.

\end{abstract}

\maketitle 

\section{Introduction} 

Mechanics is a part of every engineering curriculum, being one of the starting points in developing critical scientific thinking. Objects are first modeled as point particles, regardless of their shape and structure. Then, the model of a rigid body is introduced, which disregards deformations caused by the action of external forces \cite{Freedman,Serway}. In this model, Newton’s second law can be expressed as:


\begin{equation}
    \sum \vec{\tau} = I \vec{\alpha},
\end{equation}
 where $\vec{\tau}$ and $\vec{\alpha}$ are the torque and angular acceleration vector respectively and $I$ is the object's moment of inertia. In our experimental setup, we seek to determine the moment of inertia of a disk, around its axis of symmetry, by applying a torque that generates the disk’s rotation. The setup is described in Section 1. For a blind student, it is impossible to obtain measurements in this experiment because the measuring instruments are visual. Therefore, an adaptation of the measuring devices is necessary to allow a blind student to collect data \cite{Holt,DeBuvitz,Marinho,Goncalves1,Atkin,KuvNob,Goncalves2,Toenders}. The natural way of doing this is by translating visual signals into sound signals. The open-source platform Arduino allows using various sensors and electronic components for data acquisition and control of different experiments in physics \cite{Galeriu1, Galeriu2,Galeriu3}. Adaptations for blind students are challenging for science educators, especially in laboratories of basic physics for engineering majors \cite {Millar,Dunkerton,Parry,Azevedo,Melaku,Fantin,Zou,Lewin,Scanlon}. Usually, people in this condition decide to study humanistic careers over scientific ones due to the ease of adaptations of reading material in their studies.

Our university currently has five blind students enrolled at different levels of engineering careers. This fact has urged us to provide the appropriate environment for their studies. According to UNICEF: ``Inclusive education is the most effective way to give all children a fair chance to go to school, learn and develop the skills they need to thrive,'' and ``…Inclusive education means all children in the same classrooms, in the same schools…'' \cite{UNICEF}. Although these statements refer to children, we think they also apply to college students. In our Physics department, we have had previous experiences teaching blind students. For instance, we have adapted experiments to measure the linear expansion coefficient \cite{Lisboa1} and the distance between nodes in standing waves \cite{Lisboa2}. These adaptations have attracted the interest of blind students to apply to our university. 

In this work, we present a successful case of the measurement of the moment of inertia of a disk performed by a blind engineering student. We describe the adaptation of the process of data acquisition using low-cost equipment which can be replicated in other educational institutions.


\section{Experimental setup}

\subsection{Conventional experimental setup and analytical considerations}

\begin{figure}
    \centering
    \includegraphics[width=0.6\textwidth]{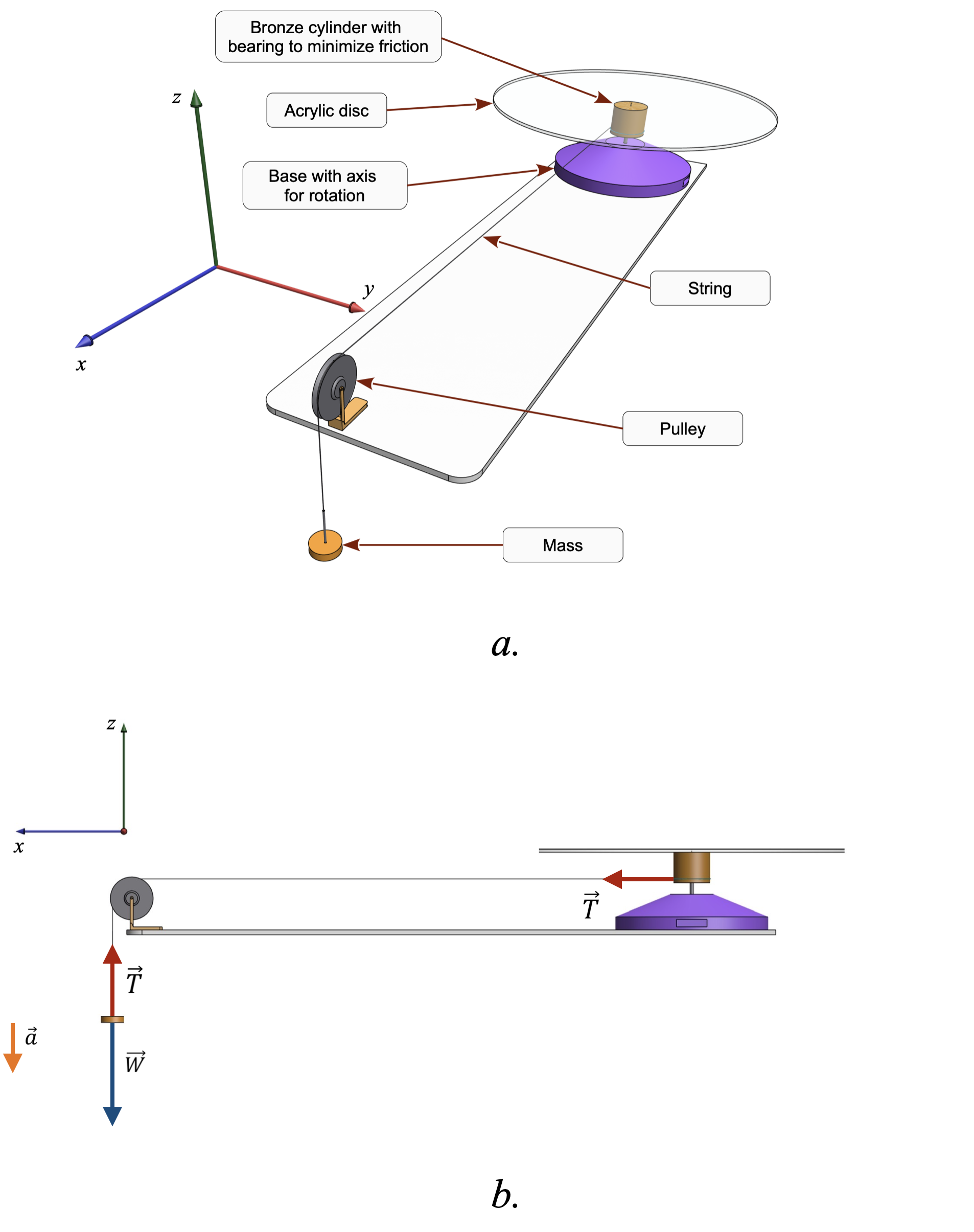}
    \caption{a. The diagram represents the experimental setup used to measure the inertia of rigid geometric objects. b. Scheme side view.}
    \label{fig1}
\end{figure}


The system consists of an acrylic disk attached to a semi-hollow bronze cylinder, which can rotate around an axis perpendicular to the disk through its center of mass. The bronze cylinder is connected to a hanging mass $m_h$ through a string and a pulley Fig.~\ref{fig1} (a) and (b). The friction in the axis and the mass of the pulley are neglected. 

The dynamical equations, considering the z-axis, vertical upwards, are:

\begin{equation}
    \textrm{Rotating system:} \, \sum \vec{\tau} = I \vec{\alpha} \rightarrow T R_{c} = I_{s} \alpha,
\end{equation}

\begin{equation}
   \textrm{Hanging mass:} \, \sum \vec{F} =  m_{h} \vec{a} \rightarrow  T - m_{h}g = - m_{h}a,
\end{equation}

where $T$ is the tension in the rope, $R_{c}$, the radius of the brass cylinder, $I_{s}$, the total moment of inertia of the cylinder-disk system, and $a$, the magnitude of the acceleration of the hanging mass. Solving the previous system of equations for the tension and using the relationship between the tangential acceleration and the angular acceleration $(\alpha = a/R_{c})$, we obtain that the expression gives the stress:

\begin{equation}
\label{tension}
    T=\left(\frac{I_{s}}{R_{c}^{2}}\right) a.
\end{equation}

Therefore, in a graph of $T$ versus $a$, the slope of the graph corresponds to the value of $I_{s}/R_{c}^{2}$ of the system cylinder-disk. The hanging mass is varied sequentially to obtain several points of the graph. The students learn how to determine the moment of inertia of simple objects by performing linear regression, which is the objective of this experiment.


\subsection{Modified experimental setup}

\subsubsection{Description of the system}



The experimental setup where the experiment was carried out is detailed in Fig.~\ref{fig2}. The information on each component is given below. We will start by describing the bronze cylinder, whose measurements are an outer diameter of 48.7 mm and a height of 39.6 mm. At the top of the cylinder, a small ``shaft" (excess of the main bronze cylinder) has been left, which helps to locate the acrylic disk concentrically. We use a bearing of 20x8 mm to minimize friction at rotational motion, with the lubricating grease replaced by a thin film of oil. In this way, when the cylinder rotates at low revolutions and under the action of small-magnitude torques, the oil allows adequate lubrication without causing considerable opposition to the system's rotation (as grease would do). 


In our adaptation, we measure angular acceleration. For this purpose, an opaque black plastic slotted disk of 100.0 mm diameter and 2.6 mm thickness attached to the base of the cylinder must be used. The disk has 100 slots of 4.0 mm long at 1.8 mm from the edge and equally spaced angularly. The slotted disk is used in conjunction with two sensors; the first sensor indicates only when the system is in rotational motion by detecting, via a photogate, slots or barriers. The motion indicator is a typical sound: \textit{beep-beep}. This sensor is used to avoid accidents when holding the disc rotating at very high speeds. 
The second sensor, a second photogate, measures the rotational acceleration of the system. This measurement is delivered to the student via an audio readout. The audio readout interface has two buttons:

\begin{itemize}
    \item B1: Performs the audio readout of the angular acceleration measurement. 
    \item B2: Resets the system to perform other measurements. 
\end{itemize}

\begin{figure}
    \centering
    \includegraphics[width=0.8\textwidth]{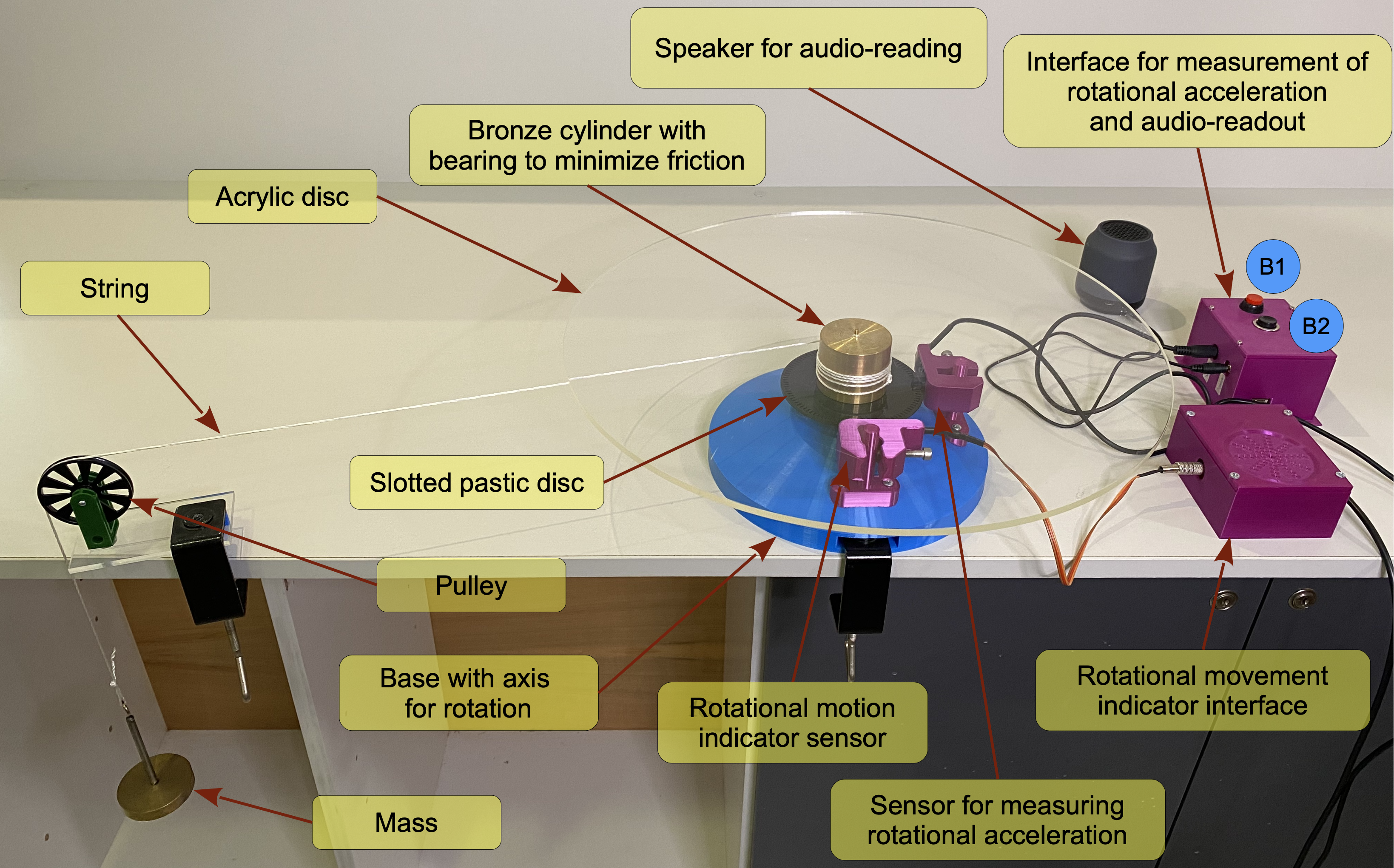}
    \caption{Scheme of the system used by a blind student to measure the moment of inertia of an acrylic disk. The image details the names of each component in addition to the control buttons B1, for the audio reading of the angular acceleration measurements, and B2, for the corresponding reset.}
    \label{fig2}
\end{figure}

The reading interfaces were manufactured on Arduino modules with the appropriate configurations and programming for each of the two sensors.

\subsubsection{Equipment calibration process}


 Traditionally, the linear acceleration of the hanging mass is measured to obtain the moment of inertia of the experiment. For this, the most common is to use a standard ultrasonic sensor placed just below the hanging mass (being careful not to hit the sensor), which provides measurements of position v/s time, velocity v/s time, and acceleration v/s time. Before the application of the modified experiment, we use this ultrasonic sensor to compare the measurements that we will obtain from the angular acceleration (in the modified experiment) and verify that they are consistent with each other, thus validating the measurements that our blind student will take in comparison to the rest of his peers during the lab session.

\begin{figure}
    \centering
    \includegraphics[width=1\textwidth]{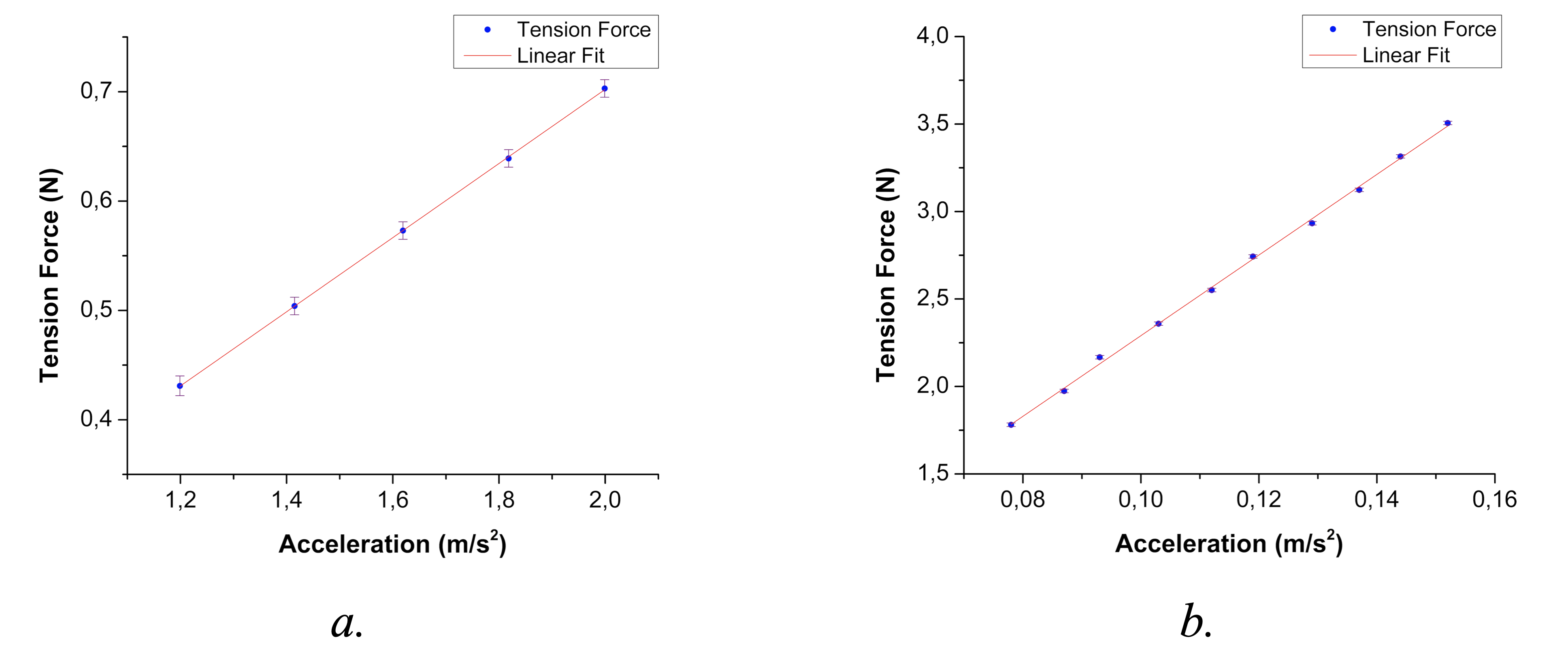}
    \caption{a. Graph of  tension v/s acceleration to obtain the rotational inertia of the bronze cylinder. b. Graph of tension v/s acceleration to obtain the inertia of the rotations of the bronze cylinder plus the acrylic disc.} 
    \label{fig3}
\end{figure}

 From the description of the equipment in Fig.~\ref{fig2}, it can be seen that this system is made up of two bodies (bronze cylinder and acrylic disc). Therefore, it is necessary first to know the rotational inertia of the bronze cylinder (including the bearing). To do this, we increase the hanging mass gradually and measure the linear acceleration for each hanging mass value without the acrylic disk. In this particular case, only five measurements could be made because the inertia of the cylinder used is tiny, which makes it challenging to apply large masses since they would generate high RMPs and the hanging mass would fall practically ``free'', making it difficult to stop it to avoid impact with the sensor. The graph of Fig.~\ref{fig3} (a) corresponds to a scatter graph of $T$ v/s $a$ for the cylinder. Applying a linear regression to the data, we obtain a slope value of $(0.339 \pm 0.002)$ kg. On the other hand, the radius of the cylinder is given by $R_{c}=0.024$ m, and using Eq.~(\ref{tension}), we find that the inertia of the disk is given by $\sim 2.01 \times 10^{-4}$  kg m$^{2}$.

Having measured the inertia of the bronze cylinder, the acrylic disc was added to the system, and the measurement procedure was repeated. Fig.~\ref{fig3} (b) shows the combined system's (disk + cylinder) scatter graph for $T$ v/s $a$. In this case, since the rotational inertia (by adding the acrylic disc) was increased considerably, we can employ larger masses, increasing the tension over the rope that causes the rotation. Consequently, it can be noted that the number of measures was also higher for this case. The RPM achieved was not that high, and the descent of the hanging mass could be controlled very well. The slope of this graph has a value of $(23.0 \pm 0.3)$ kg. Again using Eq.~(\ref{tension}), we will obtain that the inertia of the combined system is given by $\sim 1.36 \times 10^{-2}$ kg m$^{2}$.  Using then that the total measured inertia of this case can be written as $I_{s}^{me}=I_{d}^{me}+ I_{c}^{me}$, where $I_{d}^{me}$ is the inertia measured of the disk and $I_{c}^{me}$ is the inertia measured of the cylinder, we can obtain with the previous calculations that
\begin{equation}
    I_{d}^{me}=I_{s}^{me}-I_{c}^{me} \sim 1.34 \times 10^{-2} \text{kg m$^{2}$}.
\end{equation}
To compare with theoretical inertia values for homogeneous rigid objects based on their geometry, we can use the formula for disc inertia given by $I_{d}^{teo}=\frac{1}{2}M_{d}R_{d}^{2}$, where $M_{d}$ is the mass of the disc which in this case is given by $M_{d}=0.674$ kg and $R_{d}$ is the radius of the disc given by $0.2$ m. Replacing the indicated values, we get the theoretical value of inertia, the value of $1.35 \times 10^{-2}$ kg m$^{2}$. Finally, we calculate the percentage relative error given by

\begin{equation}
    \xi = \left|  \frac{I_{d}^{me}-I_{d}^{teo}}{I_{d}^{teo}} \right|,
\end{equation}
where in this case, we found by an error of $0.7$ \% is obtained. This tells us that the experience has been very well designed, and the materials and manufacturing methods have been adequate.


\subsection{How does a blind student operate and manage the equipment?}

The system's initial condition presented to the blind student consists only of the bronze cylinder with the rope completely unwound (this implies that the mass hangs freely without vertical movement). Several acrylic disks of different dimensions are presented, and the student selects one.  This selection process is necessary as we need the blind student to feel comfortable with the maneuverability of the disk that he/she will work with and carry out the experience. Consequently, the selection of the size of the disk, in this case, was with the measurements that we discussed in the previous section. Subsequently, through his sense of touch, he identifies the axis of the bronze cylinder and a small hole in the acrylic disk. This allows him to position the disk on the bronze cylinder concentrically. To prevent the cylinder from slipping in rotation concerning the disc and these rotate as a single piece, two small pieces of transparent double-contact tape are glued to the surface of the cylinder, connecting the cylinder to the disc. The student begins to rotate the disk (with one hand) slowly, and the string begins to wind around the cylinder. The motion detection sensor indicates, through a \textit{beep-beep} alarm, that the cylinder is rotating in conjunction with the disc. Using the free hand, the student ensures that the hanging mass does not hit the pulley and remains in a suitable free position to start the measurement. Once this is accomplished, press the B1 button and release the system. The motion sensor will record the angular acceleration of the disk for a time of 2 s. This experimental data acquisition time is used because the hanging mass has not yet touched the ground, even for the largest mass considered in the experiment. For the case with the bigger mass, it was measured that it would be approximately 5 cm above the ground.
Consequently, any value lower for the mass in the experience will be at a higher height at 2 s. For sequentially increasing the hanging mass, the student leaves the rope fully stretched, tactilely selects the mass to be added (of 50 grams each), and, using the rope as a guide, adds the corresponding mass. Unlike other previous experiences of laboratory work carried out with blind students, it can be noticed in our study case that it has an excellent spatial location in the environment. In conversation with the student, he tells us that this may be because he practices sports professionally, which helps him improve this aspect.


\section{Results and Discussion.}

\subsection{Description of the procedure and measurements performed by the blind student.}

\begin{table}[h!]
\centering
\caption{Data measured by the blind student}
\begin{ruledtabular}
\begin{tabular}{c c c c c p{5cm}}
Mass (kg) & Acceleration (m/$s^2$) & Angular acceleration (rad/$s^2$) & Tension (N) \\
\hline	
0,183 & 0,085 & 3,49 & 1,78 \\
0,233 & 0,102 & 4,19 & 2,26 \\
0,283 &	0,119 & 4,89 & 2,74 \\
0,333 & 0,144 & 5,93 & 3,22 \\
0,383 & 0,161 & 6,63 & 3,70 \\
0,433 & 0,180 & 7,38 & 4,17 \\
0,483 & 0,204 & 8,37 & 4,64 \\
\end{tabular}
\end{ruledtabular}
\label{data}
\end{table}

\begin{figure}
    \centering
    \includegraphics[width=0.7\textwidth]{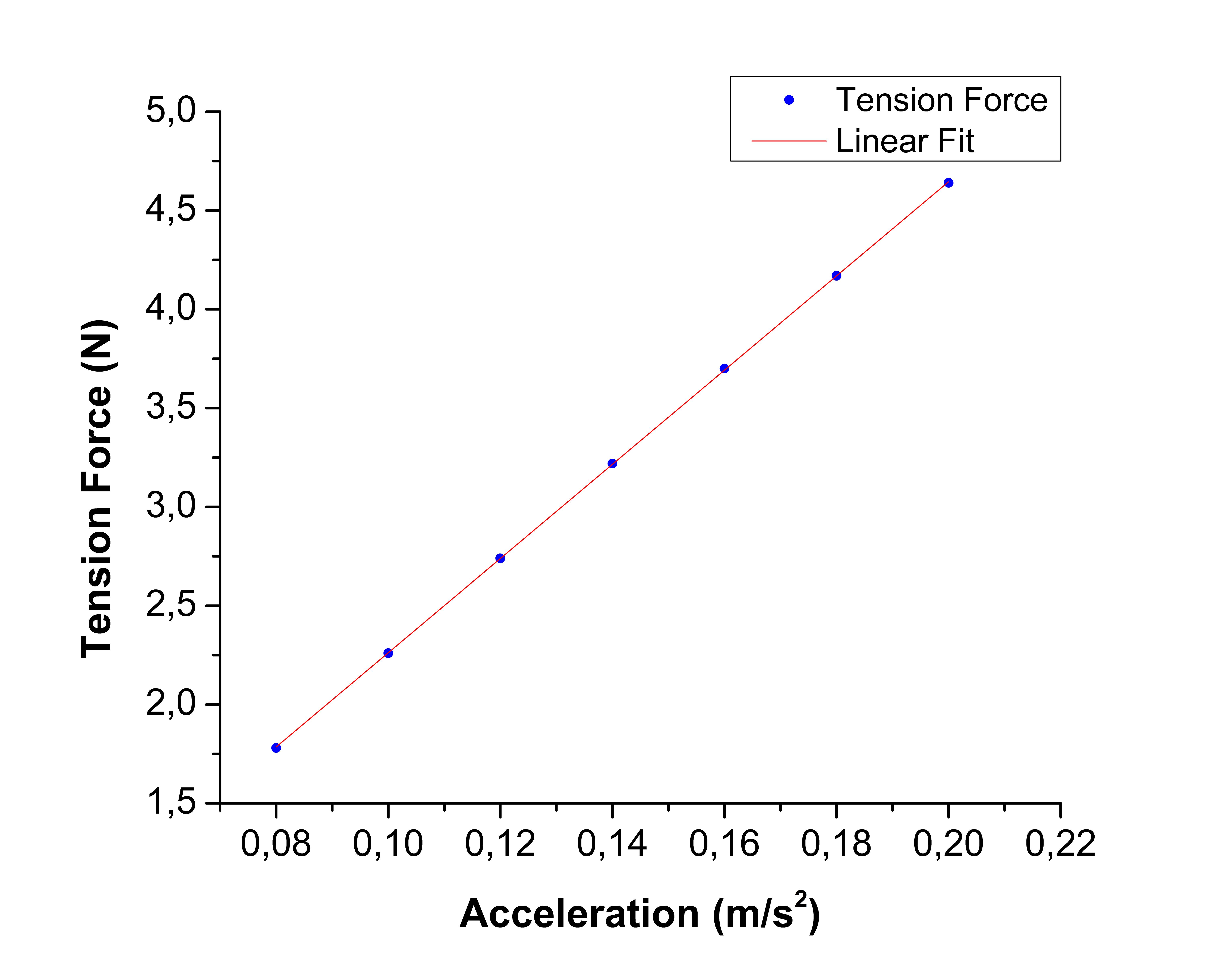}
    \caption{Graph of tension v/s acceleration corresponding to the data acquired by the blind student. In the same way, as in the calibration graphs, only the slope data will be used for the analysis. Error bars were not required of the student.}
    \label{fig4}
\end{figure}

The measurements made by the blind student can be seen in Table \ref{data} and displayed in the scatter graph of Fig.~\ref{fig4}. In this case, error bars of measurements made in a physics laboratory are not included. This is because the time required by the student to develop the experience was much greater than that used by a student without decreased vision. Consequently, the student was informed that these values would not be strictly necessary for this work. This observation helped us, as teachers, to have this consideration of extra time for future experiences to be designed on issues of educational inclusion.

The results obtained gave a slope value of $ \sim 23.84 \pm 0.04$ kg. The result of Adj. R-Square was 0.9998. The calculation of the disc's moment of inertia with the data obtained by the student gives us a value of $\sim (1.39 \times 10^{-2})$kg m$^2$. If we compare with the calculation of the moment of inertia according to the measure with the acceleration of the ultrasonic sensor, there is a percentage difference of $3.0$\%.


\section{Conclusion}

We designed, manufactured, and calibrated a physics laboratory experience to be operated by a blind student, in which the moment of inertia of an acrylic disc could be measured. From the inclusive perspective of this design, the main objective is that the student can develop the laboratory autonomously in handling equipment and instruments, controlling variables, and acquiring experimental data.

This is a traditional experience in physics laboratories for science and engineering careers. However, the readings of the physical quantities provided by the measuring instruments are mostly done visually. In this sense, we designed an instrument that would allow us to measure angular accelerations and provide the reading value aurally. To achieve this, a disk with slots was implemented near its perimeter. It was attached to the base of the bronze cylinder that rotated, practically without friction, around an axial axis.

Concentrically on the cylinder, the acrylic disk to which you want to measure the moment of inertia is located. A photo door sensor reads the slot/blocking of the slotted  disk and sends the signals to the interface that measures the angular acceleration of the rotating system. By pressing a button, the measurement is vocalized into a small speaker. Another sensor with similar characteristics was incorporated to warn with a \textit{beep-beep} alarm when the system is in rotation.

 The blind student carried out the laboratory in an autonomous way, obtaining a result of the moment of inertia of the disk that differs from the calibration data by $3.0$\%. It is worth highlighting the fact that the student was able, with his means: to locate the disc in the proper position on the bronze cylinder, carefully wind the rope preventing the mass holder from hitting the pulley, release the disc and let it rotate freely and, finally, at the indicated moment, press the buttons to carry out the measurements. Once the reading was audited and recorded on the note, the student stopped the disk and increased the size of the hanging mass by adding an extra mass. This way, he developed the entire laboratory satisfactorily and complied with the practical objectives.

An important conclusion is that this experience took longer to complete than its peers in the same lab session. Therefore, it is vital to consider extra time when conducting a self-contained inclusive lab experience for blind students. The data logger process takes longer time than we planned the first time.

\begin{acknowledgments}

A. L. acknowledges the constant support given to the realization of this practical laboratory by Daniel Andrade. F.J.P. acknowledges financial support from ANID Fondecyt, Iniciación en Investigación 2020 grant No. 11200032, ``Millennium Nucleus in NanoBioPhysics” project NNBP NCN2021 \textunderscore 021 and USM-DGIIE. Both authors are grateful for the comments made by Professor Gonzalo Fuster to improve the manuscript.

\end{acknowledgments}

\end{document}